\documentclass[prr,superscriptaddress,twocolumn]{revtex4-1}
\pdfoutput=1
\usepackage[english]{babel}
\usepackage[T1]{fontenc}
\usepackage{amsmath,amsthm,amssymb,amsfonts,graphicx}
\usepackage{xcolor}
\usepackage{tikz} 
\usetikzlibrary{positioning}
\usepackage{xcolor}
\usepackage{cancel}
\usepackage{braket}
\usepackage{calc}
\usepackage{bbold}
\usepackage{mathrsfs}
\usepackage[colorlinks,citecolor=blue]{hyperref}
\usepackage{url}
\usepackage{color}
\usepackage{float}
\usepackage{framed}
\usepackage{bbold}
\usepackage{tikz}
\usepackage{lipsum}

\definecolor{shadecolor}{gray}{.9}
\def \beq {\begin{equation}}
\def \edq {\end{equation}}
\def \bes {\begin{subequations}}
\def \eds {\end{subequations}}
\def \beqn {\begin{equation*}}
\def \edqn {\end{equation*}}

\def \dag {\dagger}

\def \up {\uparrow}
\def \down {\downarrow}

\def \sm {\sigma}

\def \tr {\text{Tr}}

\def \det {\text{det}}

\begin{document}

\title{Quantum-enhanced performance in superconducting Andreev-reflection engines}

\author{Gonzalo Manzano}
\affiliation{Institute for Cross-Disciplinary Physics and Complex Systems IFISC (UIB-CSIC), \\ E-07122 Palma de Mallorca, Spain}
\author{Rosa L\'{o}pez}
\affiliation{Institute for Cross-Disciplinary Physics and Complex Systems IFISC (UIB-CSIC), \\ E-07122 Palma de Mallorca, Spain}

\begin{abstract}
When a quantum dot is attached to a metallic reservoir and a superconducting contact Andreev processes leads to a finite subgap current at the normal lead and the creation or destruction of Cooper pairs. Andreev-reflection engines profit from the destruction of Cooper pairs to provide the work needed to set a charge current at the normal-conductor contact generating electrical power. For this power-transduction device high power and large efficiencies in quantum-mechanically enhanced regimes are demonstrated. There thermodynamic trade-off relations between power, efficiency and stability, valid for any classical engine are overcome, and kinetic constraints on the engine precision are largely surpassed in arbitrary far from equilibrium conditions.
\end{abstract}

\maketitle
\section{Introduction}
At the interface of a normal metal and a superconductor, Andreev reflection allows electron to hole conversion by means of the creation of a Cooper pair in the superconductor \cite{Andreev1964}.  The superconducting leakage in the vicinity of a normal metal induces superconducting correlations at the normal side that have strong impact in the electronics of such systems. Hybrid superconducting-normal (NS) setups have captured great interest both from the theoretical
\cite{doi:10.1080/00018732.2011.624266} and experimental sides \cite{Silvano10}. The related phenomena, such as the Josephson effect \cite{jarillo-herrero_quantum_2006,jorgensen_critical_2007, PhysRevLett.129.207701,PhysRevB.101.245427}  and multiple Andreev reflections \cite{buitelaar_multiple_2003}, are employed as quantum advantages for creating a wide range of possibilities for new electronic devices, including supercurrent transistors \cite{Baselmans99,Huang2002,van_dam_supercurrent_2006}, generators of spin-entangled electrons 
 \cite{Choi00,Recher01,Samuelsson04,PhysRevB.63.165314,Lesovik2001,Sauret2004,Hofstetter2009,Herrmann10,Wang2022}, superconducting quantum interference devices \cite{Sueur2008,Petrashov1995,Dimoulas1995,Pothier1994,doi:10.1126/science.239.4843.992,Spathis_2011}, superconducting single-photon detectors \cite{Natarajan_2012}, NS-based qubits like Andreev qubits \cite{doi:10.1126/science.1231930,PhysRevLett.115.127002,doi:10.1126/science.abf0345,doi:10.1126/science.abf0345} and topological qubits \cite{doi:10.1126/science.1222360,doi:10.1126/science.aaf3961,PhysRevX.6.031016}. The latter proposed as building blocks for fault-tolerant quantum computation.

Beyond these applications hybrid platforms are perfect candidates for performing good thermal machines \cite{PhysRevB.94.054506} being proposed in implementations of quantum  refrigerators and heat engines~\cite{Leivo_1996,Pekola_2004,doi:10.1063/1.4875917,Rafa17,Fornieri_2017,Mastomaki_2017,Rafa18,Potts19,PhysRevB.103.085409,Tabatabei22,Blasi_2023}. To progress in this field a comprehensive understanding of their quantum thermodynamics is certainly needed. For such purpose we consider a normal conductor acting as a small quantum dot (QD) that is tunnel coupled to a metallic contact and a superconducting electrode. Quantum dots are attractive for thermodynamics as they can display some quantum advantages, namely, they exhibit quantum level discretization behaving as energy filters and they can be attached to quantum materials such as reservoirs of coherent states, i.e., Cooper pairs reservoirs.

\begin{figure}[t]
\centering
\includegraphics[width=1.0\linewidth,angle=0]{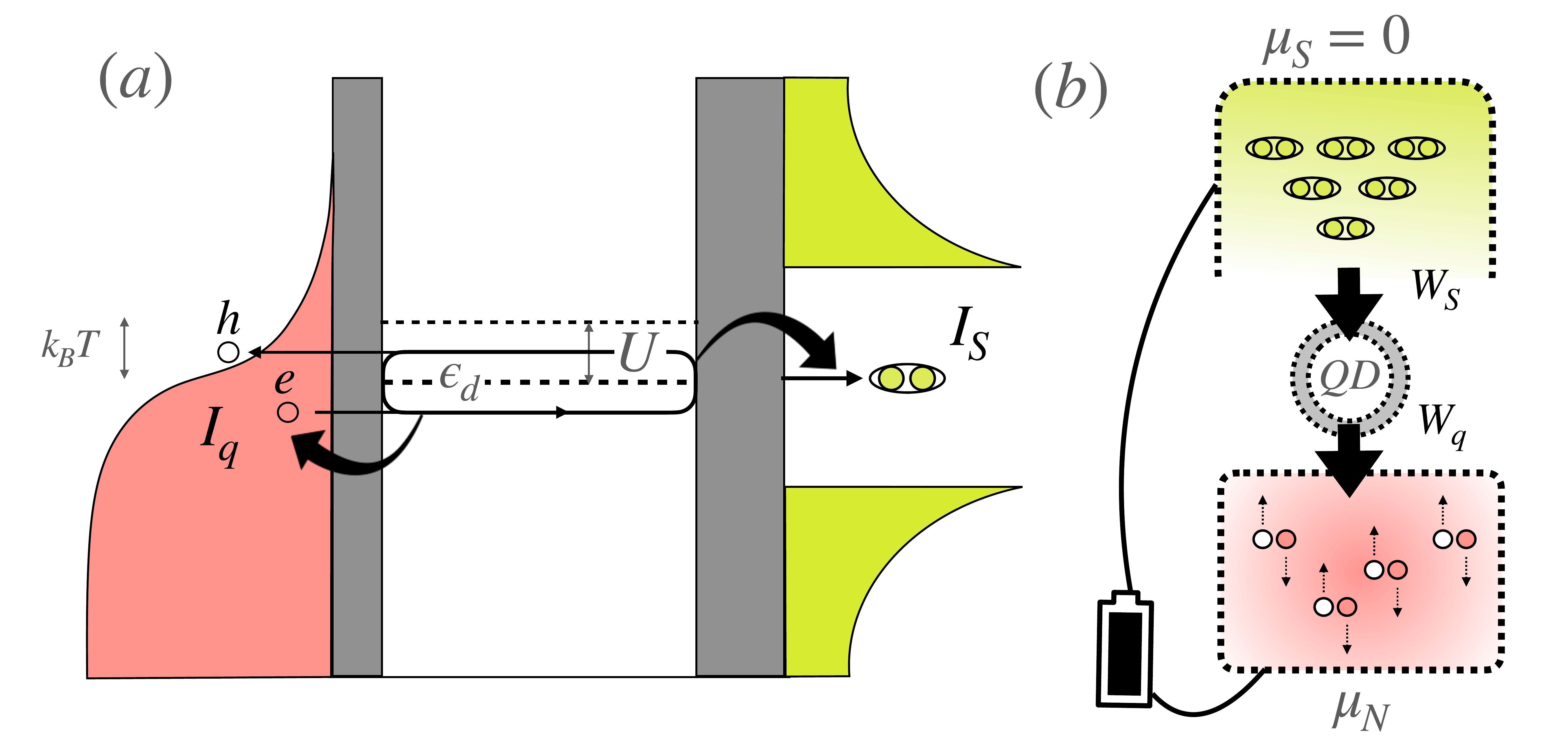}
\caption{(a) Illustration of the normal-quantum dot-superconducting engine. In Andreev reflection an electron from the normal metal (left pink terminal) is injected in the quantum dot resulting (i) a retroreflected hole at the normal reservoir and (ii) a coherent Cooper pair at the superconductor (right green terminal).
(b) The superconductor acts as a source of coherent states (Cooper pairs) that are employed through the quantum dot to generate an electrical current eventually delivering electrical power.
}
\label{fig:1}
\end{figure}

In this Letter, we characterize the quantum performance of a minimal hybrid quantum device based on Andreev-reflection processes at the level of its efficiency, power and its reliability. The device is able to operate in nonequilibrium steady-state conditions as an engine transforming coherent Cooper pairs into an electrical current (see Fig.~\ref{fig:1}). We dub this device as an Andreev-reflection engine. Contrary to standard heat engines converting heat into work~\cite{Kosloff14}, this engine is an example of a work-to-work converter, a energy transduction mechanism typical in soft nanomachines, working in the presence of various concentration gradients, external forces and torques, or electric fields~\cite{Seifert11,Proesmans16,Weiss19}. Here the destruction of Cooper pairs in a quantum dot previously loaded from a superconductor, generates an electron-hole pair at the normal contact due to the Andreev processes, i.e. a finite charge current going against the applied bias voltage ($\dot{W}_q > 0$), at expenses of a coherent input power from the superconductor ($\dot{W}_S < 0$). Inversely, because of Andreev processes, a hole can be injected into the normal contact  to create a Cooper pair at the superconducting side ($\dot{W}_S > 0$) and a retro-reflected electron at the metallic contact ($\dot{W}_q < 0$).

Remarkably, the Andreev-reflection engine, shows large violations of the so-called Thermodynamic Uncertainty Relation (TUR) and Kinetic Uncertainty Relation (KUR) that are responsible of a quantum-enhanced performance of the engine operation. The TUR is a universal nonequilibrium relation in the form of an inequality, that imposes strict bounds on the precision of generic currents through a system following classical Markovian evolution in terms of its dissipation as measured by the entropy production rate~\cite{Barato2015,Todd2016,Horowitz2020}. Relevant applications of the TUR include the estimation of dissipation in systems where only partial information is available~\cite{Seifert19}. Moreover, the TUR has been used to asses the universal trade-off between efficiency, power and their fluctuations in steady-state heat engines, predicting an inevitable drop of stability in any classical engine as their power and efficiency simultaneously increase~\cite{Pietz2018}. As a consequence, violations of the TUR have been proposed as a witness of quantum-thermodynamic signatures in the operation of quantum devices in general \cite{PhysRevLett.120.090601,Segal18,PhysRevX.11.021013}, and quantum heat engines in particular~\cite{Ptaszy2018,Liu19,Cangemi20,Kalaee21,Mitchison21,Souza22}. However, the relevance of the regimes where the TUR has been found to break down in previous works is still unclear, given also the small magnitude of the observed violations in many cases. Similarly to the TUR, the KUR consists of a universal bound on generic system currents in terms of the dynamical activity~\cite{Baiesi19}, which, contrary to dissipation, is symmetric under time-reversal~\cite{Maes2018}. Albeit passing almost unnoticed, the KUR can provide a powerful complement to the TUR in the description of engines working in nonequilibrium steady states~\cite{Potts22}, specially in far-from-equilibrium regimes~\cite{Baiesi19,Hiura21} where the TUR becomes far from tight~\cite{Horowitz2020} (see also Ref.~\cite{KTUR} for a unification of TUR and KUR). Extensions of the original TUR and KUR for quantum dynamics have been also recently reported \cite{Guarnieri19,Carollo19,PhysRevLett.126.010602,VanVu22}. As we will discuss in the following, Andreev-reflection engines are able to show unprecedentedly large violations of the KUR at maximum power, combined by notable violations of the TUR in high-power and efficiency regimes. Moreover, both inequalities can be violated simultaneously in a regime with balanced efficiency and close-to-maximum power output.

\section{Andreev-reflection engine model}
We consider a single QD device with Hamiltonian $H_d = \sum_\sigma \epsilon_\sigma  d_\sigma^\dagger d_\sigma + U d_\uparrow^\dagger d_\downarrow d_\uparrow^\dagger d_\downarrow$, where $\epsilon_\sigma$ denotes the energy level of the dot for electrons with spin $\sigma = \{ \uparrow, \downarrow \}$, and $U$ stands for the Coulomb repulsion between electrons. Eventually, one can consider the action of a magnetic field by including a Zeeman splitting in the dot level energy as $\epsilon_{\uparrow, \downarrow}=\epsilon \pm \Delta_Z$ with $\Delta_Z=\mu_B g B$ ($\mu_B$ being the Bohr magneton, $g$ the gyromagnetic factor and $B$ the applied magnetic field).
The QD is weakly coupled to both a superconducting electrode and to a metallic contact which acts as a thermal reservoir, allowing tunneling of electrons to the QD. Both terminals are kept at a constant temperature $T$ and we assume, without loss of generality, vanishing chemical potential in the superconductor, $\mu_S = 0$. We are interested in the limit of a large superconducting gap, $\Delta \rightarrow \infty$ (subgap transport)~\cite{Arovas00,Tabatabaei2020,nnano.2013.267,PhysRevB.95.180502,bargerbos2022spectroscopy,doi:10.1080/00018732.2011.624266}, that allows to approximate the effect of the superconductor on the dot as a coherent driving, described by an effective time-dependent Hamiltonian reading $H_S(t) = \Gamma_S (d_{\up}^{\dag} d_{\down}^{\dag} e^{i(2 \epsilon + U) t} +  d_{\down}d_{\up} e^{-i(2 \epsilon + U) t})$ in Schr\"odinger picture. Here $\Gamma_S$ is a local pairing term that accounts for the proximity effect of the superconductor on the QD.  The QD is tunel-contacted to a normal electrode characterized by tunneling rate $\Gamma_N$. Assuming Born-Markov approximations (either $\Gamma_N \ll |\epsilon - \mu_N|$ or $\Gamma_N \ll k_B T$~\cite{Potts_2021}) and weak pairing  ($\Gamma_S \sim \Gamma_N)$, the dynamical evolution of the system can be modeled by a local quantum master equation in Lindblad form describing the driven-dissipative dynamics of the dot (see Appendix~\ref{app:master}) reading, in the interaction (rotating) frame with respect to $H_d$:
\begin{equation}\label{eq:meq}
\dot{\rho}(t)=-i \Gamma_S [d_{\up}^{\dag} d_{\down}^{\dag} + d_{\up} d_{\down},\rho(t)] + \sum_{k} \mathcal{D}_k[\rho(t)],
\end{equation}
with dissipators $\mathcal{D}_k[\rho] := L_k \rho L_k^\dag -\frac{1}{2} \{L_k^\dag L_k \rho\}$ and a set of eight Lindblad operators labeled as $L^+_{\sigma, \delta} = \Gamma_N f(\epsilon_\sigma + \delta U) d_\sigma^\dagger$ and $L^-_{\sigma, \delta} = \Gamma_N [1 - f(\epsilon_\sigma + \delta U]) d_\sigma$, with $\delta = \{0, 1\}$ (and $k=\{\sigma,\delta\}$). These terms describe the incoherent tunneling of electrons from the normal-metal reservoir, jumping inside ($+$) or outside ($-$) the dot, at given energy. Here $\Gamma_N$ is the tunneling rate between the dot and the normal electrode, and $f(E)=1/[1-\exp[(E-\mu_N)/k_B T]$ is the Fermi-Dirac distribution with $\mu_N$ the chemical potential of the normal-metal reservoir.
From the master equation~\eqref{eq:meq} we analytically obtain the density matrix in the steady state $\rho(t \rightarrow \infty) = \pi$ ($\dot\rho =0$) from which we compute the average steady-state energy current $\langle J_E \rangle := \tr[H_d \sum_k \mathcal{D}[L_k](\pi)]$ and charge current $\langle J_q \rangle := e \sum_{\sigma} \tr[ d_\sigma^\dagger d_\sigma \sum_k \mathcal{D}[L_k](\pi)]$ entering the dot from the normal-metal reservoir. They read: 
\begin{subequations}
\begin{align}\label{xcurrents}
\langle J_E \rangle &= \Gamma_N \sum_{\sm, \delta} \epsilon_\sigma [f_{\sigma, \delta}^e(\pi_{0}+\pi_{\bar{\sigma}}) -  f^h_{\sigma, \delta} (\pi_{\sigma}+\pi_{\uparrow \downarrow})], \\ \label{xcurrents2}
\langle J_q \rangle &= \Gamma_N \sum_{\sm, \delta} e~[f_{\sigma, \delta}^e (\pi_{0}+\pi_{\bar{\sigma}}) -  f^h_{\sigma, \delta} (\pi_{\sigma}+\pi_{\uparrow \downarrow})],
\end{align}
\end{subequations}
where we denoted $f^e_{\sigma, \delta} = f(\epsilon_\sigma + \delta U)$ and $f^h_{\sigma, \delta} = 1 - f(\epsilon_\sigma + \delta U)$ the Fermi-Dirac distributions for electrons and holes, respectively, the populations of the density operator, $\pi_0 = \bra{0} \pi \ket{0}$, $\pi_\sigma = \bra{\sigma} \pi \ket{\sigma}$, $\pi_{\uparrow \downarrow}  = \bra{\uparrow \downarrow} \pi \ket{\uparrow \downarrow}$ and with $\bar{\uparrow} = \downarrow$. Explicit expressions are provided in Appendix~\ref{app:thermo}. Hereafter, we set electron charge $e=1$, Boltzmann constant $k_B=1$, and $\hbar=1$. 

Associated to the transport of energy and electrons between the dot and the normal metal reservoir, we identify the average heat current entering the system $\dot{Q} := \langle J_E \rangle - \mu_N  \langle J_q \rangle$, and the average electrical output power delivered to the normal-metal, $\dot{W}_q := - \mu_N \langle J_q \rangle$ as usual~\cite{Benenti2017}. In addition, we identify the average output power extracted into the superconductor from the time-dependent Hamiltonian $\dot{W}_S := -\tr[\dot{H}_S(t) \rho_S(t)]$, which in the interaction picture and for the steady-state regime reads:
\begin{equation} \label{eq:powersuper}
\dot{W}_S = i \Gamma_S (2\epsilon + U) ~(\pi_c -\pi_c^\ast),
\end{equation}
where $\pi_c = \bra{0} \pi \ket{\uparrow \downarrow}$. Notice that this is a coherent work contribution which only depends on the (non-zero) off-diagonal elements of $\pi$ in the $H_d$ basis, connecting the even parity QD states  by a Cooper pair. The first law in the setup hence reads $\dot{W}_S + \dot{W}_q = \dot{Q}$, ensuring the balance between output work in both terminals and input heat from the normal-metal~\footnote{Furthermore, one can verify that any output work on the superconducting contact comes from an energy current from the normal-metal, $\dot{W}_S = \langle J_E \rangle$, as expected.}

\begin{figure}[t]
\centering
\includegraphics[width=\linewidth]{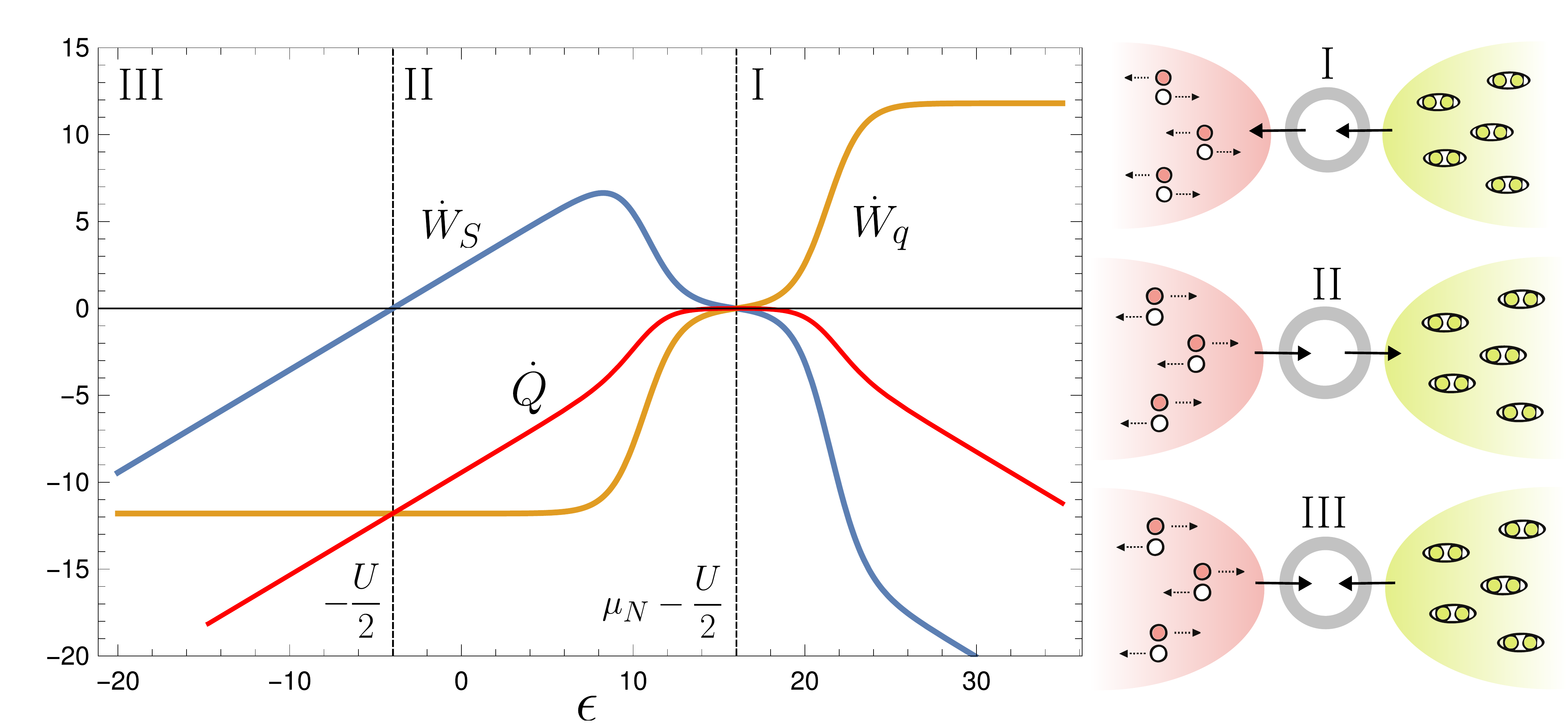}
\caption{Three operating regimes for the Andreev-reflection engine as characterized by the electric power output at the normal reservoir $\dot{W}_q$ (orange solid-line), the driving power extracted in the superconductor $\dot{W}_S$ (orange solid-line) and the heat absorbed from the reservoir $\dot{Q}$ (red solid-line) as a function of the dot energy level $\epsilon$ in $k_B T$ units (left panel) and their corresponding illustrations in terms of the direction of the currents (right panel). Power and heat current values are given in units of  $\Gamma_N k_B T$ and we took $U=8 k_B T$, $\mu_N = 20k_B T$, $\Delta_Z = 2 k_B T$,  and $\Gamma_S = 0.6 \Gamma_N$.}
\label{fig:2}
\end{figure}

 In Fig.~\ref{fig:2} we show the three main regimes of operation of the NS hybrid device as a function of the dot level $\epsilon$ for fixed values of Coulomb repulsion $U>0$ and chemical potential $\mu_N> \mu_S = 0$. In the regime (I) $\epsilon > \mu_N - U/2$, the electric current enters the normal-metal generating output power $\dot{W}_q$ at expenses of the work exerted by the superconductor, $\dot{W}_S < 0$. Maximum output electric power $\dot{W}_q^\mathrm{max} = \mu \Gamma_N[4\Gamma_S^2/(\Gamma_N^2 + 4 \Gamma_S^2)]$ is quickly reached as $\epsilon$ is increased. On the contrary if (II) $-U/2 < \epsilon < \mu_N - U/2$ work is extracted in the superconductor through the generation of Cooper pairs $\dot{W}_S > 0$ using an input electrical current from the normal-metal acting as a load, $\dot{W}_q < 0$. In this regime the superconducting power reaches it maximum, but obtaining a analytical expression for it was not possible. 
 Finally, for (III) $\epsilon < -U/2$, $\dot{W}_S < 0$ becomes negative, together with $\dot{W}_q < 0$ and hence power from both superconducting and normal-metal contacts are consumed.
 In the three regimes heat is dissipated into the environment $\dot{Q} <0$ leading to a non-negative entropy production rate: 
\begin{equation} \label{eq:sigma}
 \Sigma = - \dot{Q}/T = -(\dot{W}_S + \dot{W}_q)/T \geq 0.
\end{equation}
The equilibrium point is achieved precisely in the interface between regimes I and II, for $\epsilon = \mu_N - U/2$, where both electrical and superconducting currents change sign leading to zero dissipation. This corresponds to the condition of (global) detailed balance in the setup, obtained from $f^e_{\uparrow, 1} f^e_{\downarrow, 1} = f^h_{\uparrow, 0} f^h_{\downarrow, 0}$. 

The efficiencies of regimes I and II are obtained by dividing the output power by the corresponding input load as $\eta_I := \dot{W}_q/(-\dot{W}_S)$ and  $\eta_{II} := \dot{W}_S/(-\dot{W}_q)$, which lead to remarkably simple expressions:
\begin{eqnarray}
    \eta_I = \frac{2 \mu_N}{2\epsilon + U} \leq 1, ~~~~ \eta_{II} = \frac{2 \epsilon + U}{2 \mu_N} \leq 1,
\end{eqnarray}
where the upper bounds follow from the Second Law in the setup, as expressed in Eq.~\eqref{eq:sigma}. Notice that, contrary to the case of quantum heat engines and refrigerators, the efficiency of the Andreev-reflection engine is not bounded by the Carnot efficiency, but by $1$, as it corresponds to work-to-work transducers~\cite{Seifert11}. Moreover, similarly to other models of steady-state engines working in a continuous operation mode, maximum efficiency can be only obtained by approaching the equilibrium point, leading to vanishing power output~\cite{Kosloff14,Shiraishi16,Benenti2017}, and it verifies universal relations constraining maximum power and minimum dissipation in the linear response regime~\cite{Proesmans16,Proesmans16b,Saryal21}. 

\begin{figure*}[tbh]
\centering
\includegraphics[width=\linewidth]{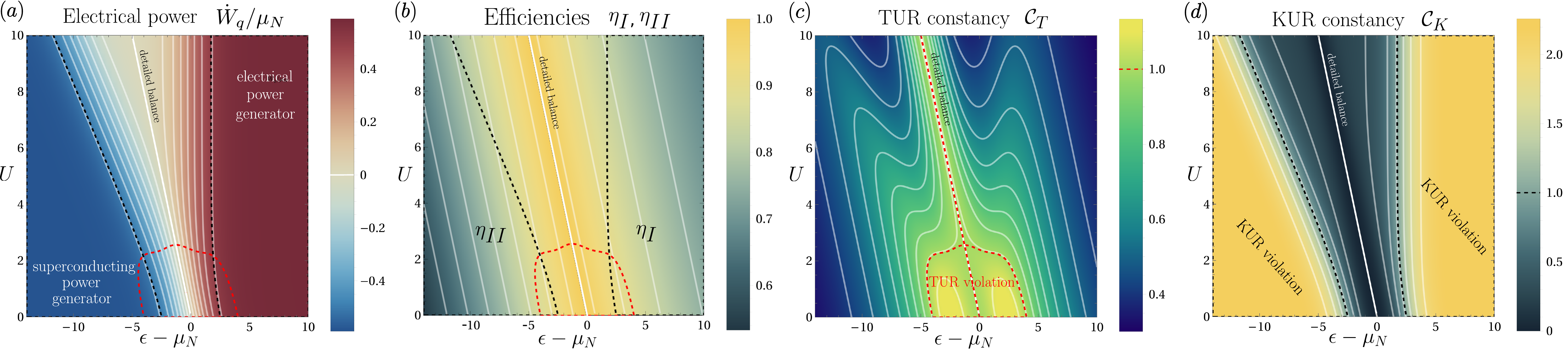}
\caption{(a) Electrical output power in $\mu_N \Gamma_N$ units versus $\epsilon - \mu_N$ (dot level height on the top the conductor chemical potential), and $U$ (Coulomb interaction). The global detailed balance condition depicted in white ($\epsilon -\mu_N = - U/2$) separates regimes with electrical power generation (regime I), and superconducting power generator (regime II). (b) Corresponding efficiency for the superconducting work to electrical power conversion ($\eta_I$) and for electrical power to superconducting work ($\eta_{II}$). (c) TUR constancy in the plane ($\epsilon - \mu_N,U$) where regions leading to TUR violations $\mathcal{C}_T>1$ are marked by the dashed red line. (d) KUR constancy in the plane ($\epsilon - \mu_N,U$) and violations $\mathcal{C}_K>1$ marked by dashed black lines. Parameters: $\Gamma_S=0.6 \Gamma_N$, $k_B T=1$ and $\Delta z = 0$}.
\label{fig:3}
\end{figure*}

\section{Enhanced stability from departures in TUR and KUR}
Beyond the power output and the efficiency of an engine, a third key element for the assessment of its performance is the stability of the output power. Generated power is affected by unavoidable fluctuations induced by environmental noise and it may spoil its reliability~\cite{Holubec18,Samuelsson19,Miller21}. The TUR has a privileged position to capture the core of the trade-off between power, efficiency and stability, since it bounds all current fluctuations by dissipation,
$\rm{Var}[X]^2/\langle X \rangle^2 \geq 2 k_B/\Sigma$, for any current, e.g. $X = J_E, J_q, ...$, where $\mathrm{Var}[X]=\sqrt{\langle (X-\langle X\rangle)\rangle^2}$ is the variance or uncertainty of the considered current~\cite{Barato2015,Todd2016,Horowitz2020}. Indeed it allows a quantitative characterization of this tradeoff through the so-called normalized constancy \cite{PhysRevLett.120.190602}, which in the present setup reads for the electric power:
\begin{eqnarray} \label{eq:TUR}
    \mathcal{C}_{T} = 2 k_B T \frac{\eta_I \mathcal{F}_q}{(1 - \eta_I) \dot{W}_q} \leq 1,
\end{eqnarray}
where $\mathcal{F}_q := \langle J_q \rangle^2 /\rm{Var}[J_q]^2$ denotes the electric  signal-to-noise ratio (squared inverse relative error) quantifying the reliability of the engine's output current, and we applied the TUR to obtain the r.h.s. inequality. This implies that low efficiency is a price to pay for accessing a large signal-to-noise ratio (and hence a reliable engine) at finite output power. We notice that the TUR strictly bounds the optimal performance of any classical steady-state engine following classical dynamics, as characterized by its power, efficiency and stability~\cite{Pietz2016}. However, in the Andreev-reflection engine, it is possible to achieve enhanced constancy in relevant operating regimes, as its quantum coherent (but Markovian) evolution induced by superconductivity allows the breakdown of the TUR. We explicitly show the TUR departure by  obtaining analytical expressions of the current variance ${\rm Var}{[{J}_q]}$ that are used to compute the signal-to-noise ratio $\mathcal{F}_q$ and the normalized constancy $\mathcal{C}_T$, using 
the full-counting-statistics formalism~\cite{Levosik96,Nazarov2003,Esposito2012,Bruderer2014}. In particular, we obtained the generalized quantum master equation including four counting fields accounting for the input-output exchange of electrons in the normal-metal contact. The cumulants are then computed following the inverse counting statistics method formulated in Ref.~\cite{Bruderer2014} from which the current variance ${\rm Var}{[{J}_q}]$ is obtained (see Appendix~\ref{app:fullcounting}).

Our results are shown in Fig.~\ref{fig:3}c where we plot $\mathcal{C}_T$ as a function of $\epsilon - \mu_N$ and the Coulomb repulsion $U$. The classical upper bound $\mathcal{C}_T=1$ is highlighted by the red dashed line, while yellow (light) tones denote quantum-enhanced constancy. While classical steady-state engines achieve unit constancy only at the equilibrium point (detailed balance line in the figure), our Andreev-reflection engine allows a quantum-mechanical enhancement of its thermodynamic performance ($\mathcal{C}_T \simeq 1.2$) in regimes of high efficiency $\eta > 0.9$ and near the maximum output power region $\dot{W}_q~\simeq (0.4$ --- $0.5) \mu_N \Gamma_N$, as can be appreciated in Figs.~\ref{fig:3}a and \ref{fig:3}b, in which the region leading to TUR violations has been also included to guide the eye.

Far from equilibrium,  however, dissipation is not the only quantity of interest,  but time-symmetric quantities may play an important role as well~\cite{Maes2018}. 
An important example is dynamical activity (or frenesy)~\cite{Roldan19,Voituriez14,Garrahan07,Volpe11} characterizing the total volume of transitions in a system per unit time, regardless of the net current directions.  In our setup it reads:
\begin{eqnarray} \label{eq:dynamical}
  \mathcal{K}= \Gamma_N \sum_{\sm, \delta} [f_{\sigma, \delta}^e(\pi_{0}+\pi_{\bar{\sigma}}) +  f^h_{\sigma, \delta} (\pi_{\sigma}+\pi_{\uparrow \downarrow})],
\end{eqnarray}
for the explicit expression see~\cite{supp}. It has been shown that dynamical activity also bounds all steady-state currents by means of the KUR, as $\rm{Var}[X]^2/\langle X \rangle^2 \geq \mathcal{K}^{-1}$  for $X = J_E, J_q, ...$, providing a tighter constraint to the noise-to-current ratio than the TUR in far for equilibrium conditions~\cite{Baiesi19,Hiura21}. 
The KUR allows us to directly introduce a new normalized constancy bounded by $1$ for classical Markov dynamics in nonequilibrium steady states:
\begin{equation} \label{eq:KUR}
  \mathcal{C}_{K}=\frac{\mathcal{F}_q}{ \mathcal{K}}  \leq 1,
\end{equation}
to be compared with the TUR-based normalized constancy in Eq.~\eqref{eq:TUR}. The above inequality provides a complementary (strict) constrain on the stability of any classical steady-state engine and highlights the fact that in order to increase signal-to-noise ratio, a larger dynamical activity is necessary, independently of the incurred dissipation.

 Remarkably, we find that the Andreev-engine exhibits large violations of the KUR-normalized constancy ( up to $\mathcal{C}_K \simeq 2.2.$) in extensive parameter regimes, as shown in Fig.~\ref{fig:3}d, as soon as we deviate from equilibrium, i.e. from the centered white line representing detailed balance conditions. The region where Eq.~\eqref{eq:KUR} is violated is highlighted in yellow (light) color, with the classical maximum value represented by the black dotted curves. This region includes parameters for which maximum power has been (almost) saturated $\dot{W}_q \simeq 0.5\mu_N\Gamma_N$ (see Fig.~\ref{fig:3}a where the region with enhanced KUR-normalized constancy is also plotted) and the engine efficiency starts to decrease with respect to the region showing TUR violations, $\eta_I \simeq 0.9 - 0.6$ (both regions are also compared in Fig.~\ref{fig:3}b). Therefore, it turns out that the most accentuated quantum consequences on the engine precision 
 occur indeed in far from equilibrium conditions, and are in general undetected by the TUR. There is, however, a small region where both TUR and KUR are simultaneously violated, where the Andreev-reflection engine shows a high output power ($\dot{W}_q \simeq 0.5 \mu_N \Gamma_N$) together with a stability not achievable with similar values of efficiency ($\eta_I \simeq 0.9$) by any classical steady-state engine.
 
\section{Experimental Directions and Closing}
Hybrid systems benefit from Andreev processes serving as the basis for the construction of Andreev-reflection engines operating as work-to-work transducers. Notably, these engines exhibit significant violations of the TUR and the KUR at large efficiencies and maximum power (doubling KUR violations predicted in normal double quantum dots \cite{Potts22}). We propose for its experimental realization a semiconductor nanowire QD as motivated by the recent experimental activity on the subgap transport regime \cite{nnano.2013.267,PhysRevB.95.180502,bargerbos2022spectroscopy}. Specifically, a device consisting of a nanowire QD  tunnel-coupled to a normal electrode, and partially covered by a layer of a large gap superconducting material, where our results may be tested from measurements of current fluctuations. (Typical parameters: $\Delta = 2.5$meV (Niobium)  \cite{Mourik12,Deng2012}  or $\Delta \sim 0.55$ (Vanadium with Ti and Al), $U\approx 0.2$meV, $T\in[0.1$K,$1$K]  \cite{nnano.2013.267}, and tunneling $\Gamma_N\in[1\mu$eV,$5\mu$eV] as in experiments in the weakly coupled regime \cite{Foxman93,PhysRevLett.123.117701}). Out of the large gap regime we expect that quasiparticle transport to weak the violation of TUR and KUR. Coherent assisted devices such as the Andreev-reflection engine presented here showing large departures from the classical TUR and KUR bounds, open the door to construct highly stable and efficient quantum machines under low-dissipation conditions. On the other hand, the ultimate precision bounds achievable in hybrid normal-superconducting quantum conductors remains an open question for future research~\cite{timpanaro2021precise}.

\acknowledgements
G. M. acknowledges funding from Spanish MICINN through the 'Ram\'on y Cajal' (RYC2021-031121-I) and 'Juan de la Cierva' programs (IJC2019-039592-I). R.L. acknowledges the financial support by the Grant No. PDR2020/12 sponsored by the Comunitat Autonoma de les Illes Balears through the Direcció General de Política Universitaria i Recerca with funds from the Tourist Stay Tax Law ITS 2017-006, the Grant No. PID2020-117347GB-I00, and the Grant No. LINKB20072 from the CSIC i-link program 2021. This work has been partially supported by the María de Maeztu project CEX2021-001164-M funded by the MCIN/AEI/10.13039/501100011033.

\appendix

\section{Master equation derivation} \label{app:master}
The Hamilonian of the QD system interacting with normal and superconducting leads reads:
\begin{align}\label{Hamtot}
\mathcal{H} =&\sum_\sigma \left[\epsilon_{\sigma} d_{\sigma}^\dagger d_{\sigma} + \frac{U}{2} d_{\sigma}^\dagger d_{\sigma} d_{\bar\sigma}^\dagger d_{\bar\sigma} \right] \nonumber \nonumber \\ &+ \sum_\sigma \sum_{k_\alpha}V_{k_\alpha }(c^\dag_{k_\alpha}d_\sm+h.c) + \sum_{k_N} \epsilon_{k_N} c_{k_N}^\dagger c_{k_N} \nonumber
\\  & +   \sum_{k_S} \epsilon_{k_S} c_{k_S}^\dagger c_{k_S}+\Delta(c_{k_S}^\dagger  c_{k_S}^\dagger +h.c)
\end{align}
As in the main text, $\epsilon_{\uparrow, \downarrow}=\epsilon \pm \Delta_Z$ is the energy level for the  dot and $U$ the Coulomb repulsion. Electrons are annihilated in the  dot by the operator $d_{\sm}$  with spin denoted by $\sigma=\{\uparrow,\downarrow\}$ ($\bar{\sigma}$ denotes the opposite spin to $\sigma$). Here $k_\alpha$ is the wavevector for the electronic states in the superconducting and metallic reservoirs, $\alpha=\{S,N\}$. $\Delta$ denotes the superconducting gap. Operators $c_{k_N}$ and $c_{k_S}$ represent the destruction operator for electrons at the normal ($N$) and superconducting ($S$) electrodes. The tunneling amplitudes between the normal-to-dot and superconducting-to-dot parts are denoted by $V_{N}\equiv V_{k_N}$ and $V_{S}\equiv V_{k_S}$, respectively, that are considered independent on $k$. The relevant QD states are denoted in short as $|\uparrow\rangle$,  $|\downarrow\rangle$ for one electron on the dot with respective spin, $|0\rangle$ for no electrons, and $|\uparrow \downarrow\rangle$, corresponding to a Cooper pair in the QD.

From Eq.~(\ref{Hamtot}) we perform two approaches, namely (i) the on-site Coulomb interaction is considered weak enough to allow a mean-field treatment, and (ii) we follow Ref.~\cite{Arovas00} and we adopt an effective Hamiltonian description {corresponding to the large gap limit (subgap transport $\Delta \gg \Gamma_N, k_B T$) by taking $\Delta\rightarrow \infty$. In that case the continuum part of the charge current due to the quasiparticle contribution is negligible~\cite{Tabatabaei2020,Tabatabei22,nnano.2013.267,PhysRevB.95.180502,bargerbos2022spectroscopy}}. Under this considerations, the QD is ``proximitized" by the superconductivity introducing an effective local pairing term denoted by $\Gamma_S$. Consequently the QD Hamiltonian is composed by a bare term $H_d=\sum_{\sm} \epsilon_{\sm}d_\sm^\dag d_\sm + U d_{\uparrow}^\dagger d_{\uparrow} d_{\downarrow}^\dagger d_{\downarrow}$ and the pairing term $H_S(t)=\Gamma_S (d_{\up}^{\dag} d_{\down}^{\dag}e^{i(2 \epsilon + U) t} +  d_{\down}d_{\up}e^{-i(2 \epsilon + U) t})$, whereas the superconducting electrode is traced out. 

In the weak coupling limit of the QD normal-metal interaction, and under Born-Markov approximations { ($V_{N}^2 \ll \max\{|\epsilon_\sigma - \mu_N|, k_B T\}$~\cite{Potts_2021})}, the driven-dissipative dynamics of the system can be described by means of a quantum master equation~\cite{Breuer2002,Wis10}. Moreover, in the limit of weak interaction with the superconductor, one can treat $H_S(t)$ as a perturbation to the bare QD Hamiltonian $H_d$, and perform the secular approximation up to leading order~\cite{Tru16} to obtain a master equation in Gorini-Kossakovsky-Sudarshan-Lindblad (GKSL) or simply Lindblad form:
\begin{align} \label{eq:QME}
\dot{\rho}_S(t) =&-i [H_d + H_S(t) + H_{LS}, \rho_S(t)] \nonumber \\ &+ \sum_{k} \mathcal{D}_k[\rho_S(t)] ~+ O(V_N^2 \Gamma_S),
\end{align}
{ which remains valid as long as $O(V_N^2 \Gamma_S) \sim O(V_N^3)$ or greater. H}ere $H_{LS}$ is the so-called Lamb-shift Hamiltonian and we obtain dissipators $\mathcal{D}_k[\rho] =  L_k \rho L_k^\dag -\frac{1}{2} \{L_k^\dag L_k \rho\}$ (with $k=\{\sigma,\delta\}$) for the set of Lindblad operators $L^+_{\sigma, \delta} = \sqrt{\Gamma_N f(\epsilon_\sigma + \delta U)} d_\sigma^\dagger$ and $L^-_{\sigma, \delta} = \sqrt{\Gamma_N [1 - f(\epsilon_\sigma + \delta U])} d_\sigma$, with $\delta = \{0, 1\}$. They produce jumps between the bare $H_d$ QD states with $f(E)=1/[1+\exp[\beta(E-\mu)]$ the Fermi-Dirac distribution function [$\beta=(k_B T)^{-1}$ with $k_B$ the Boltzmann constant and $T$ the temperature] and $\Gamma_N=\pi V_{N}^2\nu_N$ is considered as a constant ($\nu_N$ is the DOS at the normal contact). 

The terms $H_{LS}$ and $ \mathcal{D}_k[\rho_S(t)]$ above are of the order $V_N^2$, while further terms of order $O(V_N^2~\Gamma_S)$ are neglected, justifying that the presence of the superconductor driving term, $H_S(t)$ does not alter the dissipative interaction between the QD and the normal metal~\cite{Wis10}. This procedure is standard for obtaining a local master equation~\cite{Tru16}, while alternative ways have been discussed e.g. in Refs.~\cite{Hofer2017,Barra15,deChiara2018,Cattaneo2019}. Moreover, as it is customary in open quantum systems, we neglect the Lamb-shift Hamiltonian $H_{LS}$ since it only leads to a (small) renormalization of the original energies~\cite{Breuer2002,Wis10}. Taking these considerations into account, and shifting to the rotating frame with respect to $H_d$, we recover, from Eq.~\eqref{eq:QME}, the master equation \eqref{eq:meq}.

From the master equation \eqref{eq:meq} we can obtain the Pauli equations for the QD level occupations $p_i := \langle i |\rho | i \rangle$ for $i = \{0, \uparrow, \downarrow, \uparrow \downarrow \}$ and the only relevant off-diagonal term $c := \langle 0 |\rho | \uparrow \downarrow \rangle$ induced by the superconductor coherent driving. Below we display the Pauli equations in matricial form as $d\vec{p}(t)/dt =W \vec{p}(t)$ with column vector $\vec{p} = (p_0, p_\uparrow, p_\downarrow, p_{\uparrow \downarrow}, c ,c ^\ast)^T$ and rate matrix $W$: 
\begin{widetext}
\begin{equation} \label{eq:W}
\begin{pmatrix} \dot p_0 \\ \dot p_\downarrow \\  \dot p_\uparrow \\  \dot p_{\uparrow \downarrow} \\ 
 \dot c \\  \dot c^\ast     
\end{pmatrix}
=
\begin{pmatrix}
-(k_{\uparrow}^+ + k_{\downarrow}^+) & k_{\downarrow}^-& k_{\uparrow}^-  & 0   &   i \Gamma_S  & -i \Gamma_S  \\ 
k_{\downarrow}^+ &  -(k_{\downarrow}^- + r_{\uparrow}^+) & 0 &   r_{\uparrow}^-  & 0  &0 \\ 
k_{\uparrow}^+ & 0 &-(k_{\uparrow}^- + r_\downarrow^+) & r_{\downarrow}^- & 0 & 0 \\
0 & r_{\uparrow}^+  & r_{\downarrow}^+  & -(r_{\uparrow}^-+r_{\downarrow}^-)                                                                                                     &-i \Gamma_S & i \Gamma_S & 
\\ i \Gamma_S & 0   & 0   & - i \Gamma_S & -\Gamma  & 0 \\  
- i \Gamma_S  & 0 & 0	& i \Gamma_S & 0 &- \Gamma
\end{pmatrix}
\begin{pmatrix} p_0 \\ p_\downarrow \\ p_\uparrow \\ p_{\uparrow \downarrow} \\ 
c \\ c^\ast      
\end{pmatrix}
\end{equation}
\end{widetext}
where $i$ is the pure imaginary complex unit and the transition rates that appear in the rate matrix $W$ read $k_\sm^\pm :=\Gamma_N f(\epsilon_{\sm})$ and $r_\sm^\pm := \Gamma_N f(\epsilon_{d\sm} +U)$. Moreover, for the ease of notation, we used $\Gamma:= \sum_{\sm} (k_{\sm}^+ + r_{\sm}^-$)/2. Notice that the transition rates $k_{\sm}^{\pm}$ and $r_{\sm}^\pm$ stand for adding an electron on the dot ($+$) with spin  $\sm $ and for subtraction of an electron ($-$) with spin $\sm$, and we distinguished the cases where the quantum dot is either empty ($k$-rates) or singly occupied ($r$-rates).

\section{Main thermodynamic quantities} \label{app:thermo}
In this appendix we provide some of the analytical expressions for the main quantities characterizing the performance of the Andreev-reflection engine introduced in the main text. First, we provide expressions for the energy and charge currents  in Eqs.~\eqref{xcurrents} and \eqref{xcurrents2}:
\begin{align}
 \langle J_E \rangle &=  \frac{4~\Gamma_N (2\epsilon + U) \Gamma_S^2 (f^e_{\uparrow, 1} f^e_{\downarrow, 1}- f^h_{\uparrow, 0} f^h_{\downarrow, 0})}{\sum_{\sigma} (f^e_{\sigma, 1} + f^h_{\sigma, 0}) \big(4 \Gamma_S^2 + \Gamma_N^2[1 - \prod_\sigma (f^e_{\sigma, 1}- f^e_{\sigma, 0})]\big)},~~ \\
\langle J_q \rangle &= \frac{4~\Gamma_N  \Gamma_S^2 (f^e_{\uparrow, 1} f^e_{\downarrow, 1}- f^h_{\uparrow, 0} f^h_{\downarrow, 0})}{\sum_{\sigma} (f^e_{\sigma, 1} + f^h_{\sigma, 0}) \big(4 \Gamma_S^2 + \Gamma_N^2[1 - \prod_\sigma (f^e_{\sigma, 1}- f^e_{\sigma, 0})]\big)},~~
\end{align}
from which we can easily check that $ \langle J_E \rangle = (2\epsilon + U) \langle J_q \rangle$, and that the detailed balance condition, $f^e_{\uparrow, 1} f^e_{\downarrow, 1} = f^h_{\uparrow, 0} f^h_{\downarrow, 0}$ leads immediately to $\langle J_E \rangle = \langle J_q \rangle = 0$. On the other hand the average output power in the superconductor from Eq.~\eqref{eq:powersuper} reads:
\begin{eqnarray}
\dot{W}_S = \frac{4~\Gamma_N (2\epsilon + U) \Gamma_S^2 (f^e_{\uparrow, 1} f^e_{\downarrow, 1}- f^h_{\uparrow, 0} f^h_{\downarrow, 0})}{\sum_{\sigma} (f^e_{\sigma, 1} + f^h_{\sigma, 0}) \big(4 \Gamma_S^2 + \Gamma_N^2[1 - \prod_\sigma (f^e_{\sigma, 1}- f^e_{\sigma, 0})]\big)},~~~
\end{eqnarray}
hence verifying $\dot{W}_S =  \langle J_E \rangle$, as pointed in the main text. The entropy production in Eq.~\eqref{eq:sigma} is then given by:
\begin{eqnarray}
    \Sigma = \frac{4~\Gamma_N ( 2 \mu - 2\epsilon - U) \Gamma_S^2 (f^e_{\uparrow, 1} f^e_{\downarrow, 1}- f^h_{\uparrow, 0} f^h_{\downarrow, 0})}{T \sum_{\sigma} (f^e_{\sigma, 1} + f^h_{\sigma, 0}) \big(4 \Gamma_S^2 + \Gamma_N^2[1 - \prod_\sigma (f^e_{\sigma, 1}- f^e_{\sigma, 0})]\big)},~~~
\end{eqnarray}
and the dynamical activity in Eq.~\eqref{eq:dynamical} reads:
\begin{eqnarray}
    \mathcal{K} = \Gamma_N [\frac{\Gamma_N^2 (K_1 + K_2) \sum_{\sigma} (f^e_{\sigma, 1} + f^h_{\sigma, 0})}{\sum_{\sigma} (f^e_{\sigma, 1} + f^h_{\sigma, 0}) \big(4 \Gamma_S^2 + \Gamma_N^2[1 - \prod_\sigma (f^e_{\sigma, 1}- f^e_{\sigma, 0})]\big)} \nonumber \\ - \frac{4~ \Gamma_S^2 K_3 (f^e_{\downarrow, 1} + f^h_{\uparrow, 0} )}{\sum_{\sigma} (f^e_{\sigma, 1} + f^h_{\sigma, 0}) \big(4 \Gamma_S^2 + \Gamma_N^2[1 - \prod_\sigma (f^e_{\sigma, 1}- f^e_{\sigma, 0})]\big)}] ~~~.
\end{eqnarray}
Finally the electric power signal-to-noise ratio appearing in TUR and KUR expressions [Eqs.~\eqref{eq:TUR} and \eqref{eq:KUR}] is given by:
\begin{align} \label{eq:signal-to-noise}
    \mathcal{F}_q = \frac{(f^e_{\uparrow, 1} f^e_{\downarrow, 1} - f^h_{\uparrow, 0} f^h_{\downarrow, 0} ) \sum_\sigma (f^e_{\sigma, 1} + f^h_{\sigma, 0})}{2 \mu \left[ F_1 + F_2 + F_3 + F_4 \right]}.
\end{align}
In the above expreesions for the dynamical activity and electric signal-to-noise ratio we have introduced the following functions to simplify the expressions:
\begin{widetext}
\begin{align}
    K_1 &= (f^{e}_{\downarrow, 0})^2 \left(f^{e}_{\uparrow, 1} f^{e}_{\downarrow, 1} + f^{e}_{\uparrow, 0} \sum_\sigma f^{h}_{\sigma, 1} - f^{h}_{\uparrow, 1} -1 \right) + f^{e}_{\uparrow, 0}\left(2 f^{h}_{\uparrow, 0} - f^{e}_{\downarrow, 1} \left[(f^{e}_{\uparrow, 1})^2 + f^{e}_{\uparrow, 1} f^e_{\downarrow, 1} - \sum_\sigma f^h_{\sigma, 1} - f^e_{\uparrow,0}(f^e_{\uparrow, 1} + 1)\right] \right) \nonumber \\
    K_2 &= f^e_{\downarrow, 0} \left( 2 + (f^e_{\uparrow, 0})^2 \sum_{\sigma} f^h_{\sigma, 1} - f^e_{\uparrow, 0} \left[ (f^e_{\downarrow, 1})^2 + f^e_{\uparrow, 1} f^e_{\downarrow, 1} - \sum_\sigma f^h_{\sigma, 1} \right] \right) + 
   f^e_{\uparrow, 0} \left[(f^e_{\downarrow,1})^2 + (f^e_{\uparrow,1})^2 + 4 (f^e_{\uparrow, 1} f^e_{\downarrow, 1} - 1)  - \sum_\sigma f^e_{\sigma, 1} \right] \nonumber \\
    K_3 &= (f^{e}_{\downarrow, 0})^2 + (f^e_{\uparrow, 1} + 1)(f^h_{\uparrow, 1} + f^h_{\downarrow, 1} + f^h_{\uparrow, 0} ) + f^e_{\downarrow, 0}(2 f^e_{\uparrow,1} - f^e_{\downarrow, 0} - f^h_{\downarrow, 1}). \nonumber \\
    F_1 &= \sum_\sigma (f^e_{\sigma, 1} + f^h_{\sigma, 0})^2 \left[f^e_{\uparrow, 1} f^e_{\downarrow, 1} - f^h_{\downarrow, 0} - f^e_{\uparrow, 1} f^e_{\downarrow, 0} (1 + f^e_{\downarrow, 1}) + f^e_{\uparrow, 0} f^e_{\downarrow, 0} (f^e_{\uparrow, 1} - f^e_{\downarrow, 1}) - 
 f^e_{\uparrow, 0} (f^e_{\uparrow, 1} f^e_{\downarrow, 1} - f^h_{\downarrow, 1}) \right], \nonumber \\
    F_2 &= 4 \Gamma_S^2 \frac{(f^e_{\uparrow, 1} f^e_{\downarrow, 1} - f^h_{\uparrow, 0} f^h_{\downarrow, 0})[8 + (f^e_{\downarrow, 0})^2 - (f^e_{\downarrow, 1})^2 + f^e_{\uparrow, 0}(f^e_{\uparrow, 0}- 8)  + f^e_{\downarrow, 0} (6 f^e_{\uparrow, 0} - 8) - 6 f^e_{\uparrow, 1} f^e_{\downarrow, 1} - (f^e_{\uparrow, 1})^2]}{4 \Gamma_S^2 + \Gamma_N^2 [1 - \prod_\sigma (f^e_{\sigma, 1}- f^e_{\sigma, 0})]}, \nonumber \\
    F_3 &= 4 \Gamma_S^2 \left(\frac{f^e_{\uparrow, 1} f^e_{\downarrow, 1} - f^h_{\uparrow, 0} f^h_{\downarrow, 0}}{4 \Gamma_S^2 + \Gamma_N^2 [1 - \prod_\sigma (f^e_{\sigma, 1}- f^e_{\sigma, 0})]}\right)^2, \nonumber \\
    F_4 &= 4 \Gamma_S^2 [10 - \sum_\sigma (f^e_{\sigma, 1} - f^e_{\sigma, 0})] + \big[18 + 5 \sum_\sigma (f^e_{\sigma, 1}- f^e_{\sigma, 0} ) + (f^e_{\downarrow, 0})^2 [f^e_{\uparrow, 0} - f^e_{\uparrow, 1}]  + f^e_{\downarrow, 1}(f^e_{\uparrow, 0} - f^e_{\uparrow, 1}) [10 + f^e_{\downarrow, 1} - f^e_{\uparrow, 0} + f^e_{\uparrow, 1}] \nonumber \\ 
    &~~~+  f^e_{\downarrow, 0} [(f^e_{\uparrow, 0})^2 + 2 f^e_{\uparrow, 1}(5 + f^e_{\downarrow, 1}) + (f^e_{\uparrow, 1})^2 - 2 f^e_{\uparrow, 0} (5 + f^e_{\uparrow, 1} + f^e_{\downarrow, 1})] \big].
\end{align}
\end{widetext}

\section{Full Counting Statistics}
\label{app:fullcounting}
In order to calculate the variance of the currents in the QD systems and test TUR and KUR expressions we employ the Full Counting Statistics (FCS) formalism \cite{Lesovik2001,Nazarov2003,Esposito2012,Bruderer2014}. A generalized quantum master equation can be obtained by including counting fields for the electron transport within dissipative terms:
\begin{align}
   \dot{\rho}_G(t, \chi_\uparrow, \chi_\downarrow) =& -i \Gamma_S [d_{\up}^{\dag} d_{\down}^{\dag} + d_{\up} d_{\down},\rho_G(t,\chi_\uparrow, \chi_\downarrow)] \nonumber \\ 
   &+ \sum_{k} \bar{\mathcal{D}}_k[\rho_G(t,\chi_\uparrow, \chi_\downarrow)],
\end{align}
where $\rho_G(t, 0, 0) = \rho(t)$ and we introduced modified dissipators (again $k= \{\sigma, \delta \}$):
\begin{eqnarray}
\bar{\mathcal{D}}_k[\rho_G] =  L_k \rho_G L_k^\dag e^{-i \chi_\sigma} -\frac{1}{2} \{L_k^\dag L_k \rho_G\}.
\end{eqnarray}
The variables $\chi_\uparrow$ and $\chi_\downarrow$ represents the counting fields for $\uparrow$  and $\downarrow$ electrons exchanged with the normal metal, respectively, associated to projective measurements of the number of particles in the normal metal. The solution of the generalized master equation above provides direct information of the generating function  for particle exchange statistics with the reservoir with different spins, $N_\uparrow$ and $N_\downarrow$, as $\mathrm{Tr}[{\rho}_G(t, \chi_\uparrow, \chi_\downarrow)] = G(\chi_\uparrow, \chi_\downarrow,t) = \int dN_\uparrow dN_\downarrow P(N_\uparrow, N_\downarrow) e^{i (\chi_\uparrow N_\uparrow + \chi_\downarrow N_\downarrow)}$.

As for the case of the original master equation, the counting-field-dependent generalized master equation can be linearized and written in matrix form as $d\vec{p}_G(t)/dt =W_G(\chi_\uparrow, \chi_\downarrow) \vec{p}_G(t)$ with: 
\begin{widetext}
\begin{equation} \label{eq:WG}
W_G(\chi_\uparrow, \chi_\downarrow)=
\begin{pmatrix}
-(k_{\uparrow}^+ +k_{\downarrow}^+) & k_{\downarrow}^- \exp{(-i\chi_{\downarrow})}& k_{\uparrow}^- \exp{(-i \chi_{\uparrow})}  & 0   &   i \Gamma_S  & - i \Gamma_S   \\ 
k_{\downarrow}^+\exp{(i \chi_{\downarrow})} &  -(k_{\downarrow}^- + r_{\uparrow}^+) & 0 &   r_{\uparrow}^- \exp{(-i \chi_{\uparrow})}  & 0  &0 \\ 
k_{\uparrow}^+\exp{(i \chi_{\uparrow})} & 0 &-(k_{\downarrow}^-+r_\uparrow^+) & r_{\downarrow}^-\exp{(-i\chi_{\downarrow})} & 0 & 0 \\
0 & r_{\uparrow}^+  \exp{(i \chi_{\uparrow})} & r_{\downarrow}^+ \exp{(i \chi_{\downarrow})} & -(r_{\uparrow}^-+ r_{\downarrow}^-)                                                                                                     &-i \Gamma_S & i \Gamma_S & 
\\ i \Gamma_S & 0   & 0   & - i \Gamma_S & - \Gamma  & 0 \\  
- i \Gamma_S  & 0 & 0	& i \Gamma_S & 0 &- \Gamma
\end{pmatrix}
\end{equation}
\end{widetext}
to be compared with the rate matrix $W$ in Eq.~\eqref{eq:W} above. The cumulants are computed following the inverse counting statistics method formulated in Ref.~\cite{Bruderer2014}. In this method the characteristic polynomial of  matrix $W_G$, namely $P(\lambda) := - \det[W_G(\chi_\uparrow, \chi_\downarrow) - \lambda \mathbb{1}]$, is written as a series in terms of powers of its eigenvalues,
\begin{align}
P(\lambda)=& \sum_{\mu = 0}^M a_{\mu} \lambda^\mu = \lambda^M + a_{M-1} \lambda^{M-1} \nonumber \\ 
&+ \sum_{\mu}^{M-2}\sum_{j=0}^{\infty} \sum_{l=0}^{\infty} a_{ \mu}^{(j, l)}\frac{\chi_\uparrow^j \chi_\downarrow^l}{j!~l!}\lambda^\mu,
\end{align}
with $M=6$ the dimension of $W_G$, and where in the second line the coefficients $a_\mu(\chi_\uparrow, \chi_\downarrow)$ have been expanded in a (two-variables) Taylor series around the vicinity of $(\chi_\uparrow = 0 , \chi_\downarrow = 0)$. Above we used the short-hand notation $a_{\mu}^{(j, l)}:= \partial_{\chi_\uparrow}^j \partial_{\chi_\downarrow}^l a_\mu|_{(\chi_\uparrow, \chi_\downarrow) = 0}$ and $a_{\mu}^{(0)} = a_\mu|_{(\chi_\uparrow, \chi_\downarrow) = 0}$. Following Ref.~\cite{Bruderer2014}, the derivatives of the polynomial coefficients with respect to the counting fields can be identified with the different cumulants of the electron currents. Therefore by knowing the series coefficients of the characteristic polynomial, the different cumulants can be obtained in a systematic way. 

The characteristic polynomial $P(\lambda)$ can be analytically obtained from $W_G$ in Eq.~\eqref{eq:WG}, from which the relevant coefficients can be calculated as:
\begin{eqnarray}
a_{0}=P(\lambda=0), ~~  a_{1} =\frac{\partial P(\lambda)}{\partial\lambda}\Biggr|_{\lambda = 0}, ~ a_{2}= \frac{1}{2} \frac{\partial^2 P(\lambda)}{\partial \lambda^2}\Biggr|_{\lambda = 0} ~~
\end{eqnarray}
As a second step, we calculated their derivatives with respect to the counting fields:
\begin{eqnarray}
a_{0}^{(j, l)}=(-i)^{j+l}~\frac{\partial^{j + l} a_0}{\partial \chi_{\uparrow}^j \partial \chi_{\downarrow}^l} \Biggr|_{(\chi_\uparrow, \chi_\downarrow) = 0}, \nonumber \\
a_{1, \sigma}^{(j, l)}=(-i)^{j+l}~\frac{\partial^{j+l} a_1}{\partial \chi_{\uparrow}^j \partial \chi_{\downarrow}^l} \Biggr|_{(\chi_\uparrow, \chi_\downarrow) = 0},  \nonumber \\ 
a_{2, \sigma}^{(j, l)}=(-i)^{j+l}~\frac{\partial^{j+l} a_2}{\partial \chi_{\uparrow}^j \partial \chi_{\downarrow}^l}\Biggr|_{(\chi_\uparrow, \chi_\downarrow) = 0}.
\end{eqnarray}

Finally, from the coefficients above we obtain the first and second cumulants of the spin charge currents, namely the average currents for spin  $\sigma = \uparrow, \downarrow$, their corresponding variances, and the covariance between the two currents:
\begin{eqnarray}
&\langle J_q^{(\uparrow)} \rangle  = - \frac{a^{(1,0)}_{0}}{a_{1}^{(0)}}, ~~ \langle J_q^{(\downarrow)} \rangle  = - \frac{a^{(0,1)}_{0}}{a_{1}^{(0)}},& \\  
&\mathrm{Var}[J_q^{(\uparrow)}] = - \frac{a_{0}^{(2,0)}+ 2 a^{(1),0}_{0} J_q^{(\uparrow)} + 2 a_{2}^{(0)} J_q^{(\uparrow) 2}}{a_{1}^{(0)}},& \\
&\mathrm{Var}[J_q^{(\downarrow)}] = - \frac{a_{0}^{(0,2)}+ 2 a^{(0,1)}_{0} J_q^{(\downarrow)} + 2 a_{2}^{(0)} J_q^{(\downarrow) 2}}{a_{1}^{(0)}},& \\
&\mathrm{Covar}[J_q^{(\uparrow)}, J_q^{(\downarrow)}] = - \frac{a_{0}^{(1,1)}+  a^{(1,0)}J_q^{(\uparrow)} + a^{(0,1)}_{0} J_q^{(\downarrow)}}{a_{1}^{(0)}}& \nonumber \\ 
& ~~~~~~~~~~~~~~ + \frac{a_{2}^{(0)} (J_q^{(\uparrow) 2} + J_q^{(\downarrow)2})}{a_{1}^{(0)}}.&
\end{eqnarray}
We emphasize that the obtained expressions for the average charge current $\langle J_q \rangle = \sum_\sigma \langle J_q^{(\sigma)} \rangle$ coincide with the expression given in Eq.~\eqref{xcurrents2}. Moreover, we point out a typo in the expression of the second cumulant $c_2$ in Eq.(11) of Ref.~\cite{Bruderer2014}, where $a_0'$ should read $a_0'' \neq a_0'$, and corrected it in the above expressions. From the variance of the electron currents with spins, $\sigma = \{\uparrow, \downarrow \}$, we can obtain the variance of the electrical power output as:
\begin{align}
    &\mathrm{Var}[W_q] = \mu_N^2 (\mathrm{Var}[J_q^{(\uparrow)}] + \mathrm{Var}[J_q^{(\downarrow)}] \nonumber \\ &~~~~~~~~~~~+ 2~ \mathrm{Covar}[J_q^{(\uparrow)}, J_q^{(\downarrow)}])  = 4 \mu_N^2  \mathrm{Var}[J_q^{(\sigma)}],
\end{align}
where the last equality follows from the fact that the symmetry of our system leads to $\mathrm{Var}[J_q^{(\uparrow)}] = \mathrm{Var}[J_q^{(\downarrow)}] =  \mathrm{Covar}[J_q^{(\uparrow)}, J_q^{(\downarrow)}]$. These expressions have been used to calculate the electric power signal-to-noise ratio given in Eq.~\eqref{eq:signal-to-noise}.

\bibliography{NS_heatengine}

\end{document}